\documentclass[11pt,twoside,center]{article}

\usepackage{graphicx}

\input{spncci-busteni17.sty}

\title{\bf SYMPLECTIC NO-CORE CONFIGURATION INTERACTION FRAMEWORK FOR \textit{AB INITIO} NUCLEAR STRUCTURE}

\author{Anna E.\ McCOY\footnote{
Department of Physics, University of Notre Dame, Notre Dame, Indiana 46556-5670, USA;
TRIUMF, Vancouver, British Columbia, V6T 2A3, Canada
} 
, Mark A. CAPRIO\footnote{
Department of Physics, University of Notre Dame, Notre Dame, Indiana 46556-5670, USA
}
, Tom\'{a}\v{s} DYTRYCH\footnote{
Nuclear Physics Institute, Academy of Sciences of the Czech Republic, 250\,68 \v{R}e\v{z}, Czech Republic}}
\date{(\today)}  
\begin{document}

\maketitle

\pagestyle{myheadings}
\markboth{A.~E.~\textbf{McCoy}, M.~A.~\textbf{Caprio}, T.~\textbf{Dytrych}
}{Symplectic No-Core Configuration Interaction Framework}

\bigskip

\begin{abstract}\it
We introduce a symplectic no-core configuration interaction (SpNCCI) framework
for \textit{ab initio} nuclear structure calculations, in a correlated many-body
basis which encodes an approximate $\grpsptr$ symmetry of the nucleus.  Such a
scheme potentially provides a means of restricting the many-body space to
include only those highly-excited configurations which dominantly contribute to
the nuclear wave function.  We examine the symplectic symmetry structure arising
in an illustrative \textit{ab initio} SpNCCI calculation for $\isotope[6]{Li}$.
We observe both the dominance of symplectic symmetry in individual wave
functions and the emergence of families of states related by symplectic
symmetry.
\end{abstract}

{\bf Keywords:} \textit{Ab initio} nuclear theory, symplectic group [$\grpsptr$], no-core
configuration interaction (NCCI).

\clearpage

\section{Introduction}
\setcounter{equation}{0}

A long-standing goal of nuclear physics is to quantitatively predict the
structure of nuclei and understand their excitation modes \textit{ab initio},
\textit{i.e.}, directly from realistic internucleon interactions. However, in
a traditional oscillator-basis no-core configuration interaction (NCCI)
calculation~\cite{navratil2000:12c-ab-initio,barrett2013:ncsm}, the dimension of
the many-body basis explodes as the number of nucleons and included
single-particle excitations is increased.  The basis
size which would be required in order to obtain quantitatively accurate
predictions for nuclei with more than just a few nucleons becomes prohibitively
large~\cite{maris2009:ncfc,maris2013:ncsm-pshell}.

The symplectic no-core configuration interaction (SpNCCI) framework introduces a
correlated many-body basis, one which encodes an approximate $\grpsptr$ symmetry
of the nucleus.  Our aims in pursuing symplectic many-body symmetry are twofold: (1)~to use
symmetries to accelerate convergence of \textit{ab initio} results and (2)~to
understand the symmetries underlying many-body correlations in nuclei, including
emergent rotation~\cite{maris2015:berotor2}.

The symplectic group in three dimensions $\grpsptr$~\cite{wybourne1974:groups}
enters into the nuclear many-body problem both through its relation to
kinematics (coordinates and momenta) and through its connection to the harmonic
oscillator.  It is generated by the bilinears in coordinates and momenta,
\textit{i.e.}, operators of the form $x_ix_j$, $x_ip_j$, $p_ix_j$, and $p_ip_j$
($i,j=1,2,3$).  It thus arises naturally in problems involving the coordinates
and momenta.  Symplectic symmetry has a close relation to collective deformation
and rotational degrees of freedom~\cite{rowe2016:micsmacs}, through its Elliott
$\grpsu{3}$~\cite{elliott1958:su3-part1,harvey1968:su3-shell} (and rotor
model~\cite{carvalho1986:sp-shell-collective}) subgroup, which is generated by the
orbital angular momentum and quadrupole operators.  The group $\grpsptr$ is also
the dynamical group for the harmonic oscillator, which defines the underlying
basis for nuclear configuration interaction (or interacting shell model) calculations.

These connections permit the construction of the symplectic shell model
formalism, originally proposed by Rosensteel and
Rowe~\cite{rosensteel1980:sp6r-shell}.  While the goal of ``diagonalising a
realistic many-nucleon Hamiltonian in a $\grpsptr\supset\grpsu{3}$ shell model
basis, to obtain a fully microscopic description of collective states from first
principles'' was envisioned early on~\cite{rowe1985:micro-collective-sp6r}, only
recently has sufficient progress been made in the computational framework for
\textit{ab initio} calculations to allow this goal to be
revisited~\cite{dytrych2007:sp-ncsm-evidence,dytrych2007:sp-ncsm-dominance,dytrych2008:sp-ncsm-deformation,dytrych2008:sp-ncsm}.
\begin{figure}[t]
\begin{center}
%
\begin{minipage}[b]{0.70\hsize}
\includegraphics[width=0.52\hsize]{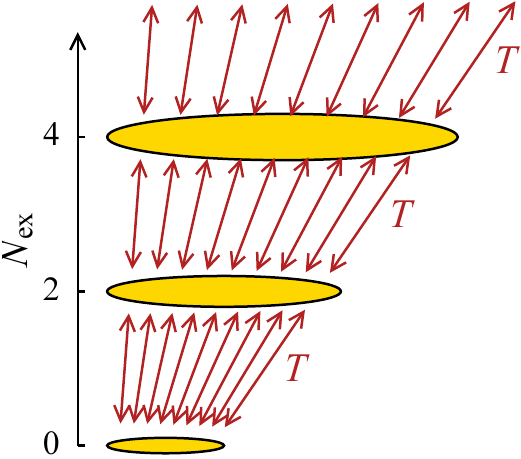}
\includegraphics[width=0.47\hsize]{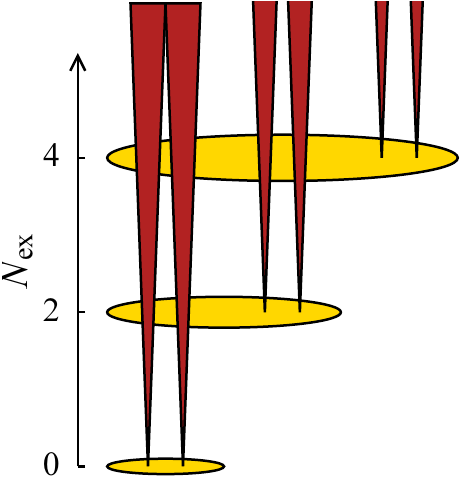}
\end{minipage}
\hspace{-0.08\hsize}
\includegraphics[width=0.3\hsize]{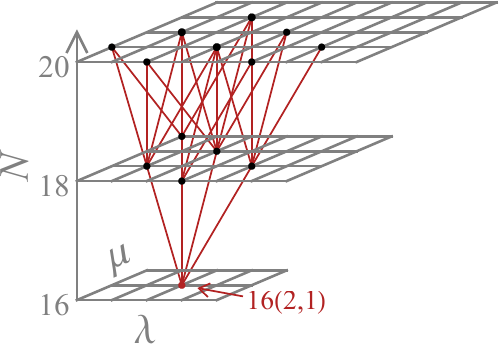}
\caption{Reorganization of the NCCI many-body space into symplectic irreps:
  (left)~the kinetic energy connects many-body states related by $0$ or $\pm2$
  oscillator quanta (as indicated by arrows), but (right)~only connects states
  within a symplectic irrep (shaded cones).  The detailed structure of an
  $\grpsptr$ irrep in terms of $\grpu{3}$ irreps $N(\lambda,\mu)$ is illustrated
  in the inset (bottom right).}
\label{fig-spncci-reorganization}
\end{center}
\figurecompensation
\end{figure}

From the viewpoint of accelerating convergence in NCCI calculations, it is
perhaps most important to note that the need for including highly excited
configurations in NCCI calculations exists, in large part, because the kinetic
energy induces strong coupling across shells. The kinetic energy operator, as a
generator of $\grpsptr$, conserves $\grpsptr$ symmetry.  Thus, by reorganizing
the many-body space according to irreducible representations (irreps) of
$\grpsptr$, it is broken into subspaces which cannot be connected by the kinetic
part of the Hamiltonian, as portrayed in Fig.~\ref{fig-spncci-reorganization}.  Combining symplectic
symmetry with the no-core configuration interaction framework potentially
provides a means of identifying and restricting the basis to include only those
highly excited configurations which dominantly contribute to the nuclear
wavefunction, thereby (it is hoped) reducing the size of the basis necessary to
obtain accurate results.

In this contribution, we briefly present a framework for \textit{ab initio}
calculations of nuclear structure in a SpNCCI scheme (Sec.~\ref{sec-framework}).
We then examine the symmetries arising in a SpNCCI calculation for
$\isotope[6]{Li}$ (Sec.~\ref{sec-symmetry}).  We observe both the dominance of
symplectic symmetry in the individual wave functions and the emergence of
families of states related by symplectic symmetry.  These results are from work
described in Ref.~\cite{mccoy2018:diss}.

\section{Symplectic NCCI framework}
\label{sec-framework}
\setcounter{equation}{0}

The SpNCCI basis for the nuclear
many-body problem consists of states which reduce an
$\grpsptr\supset\grpu{3}\supset\grpso{3}$ subgroup chain, \textit{i.e.}, which
are arranged into nested irreps of all three of these groups.  A symplectic
irrep is built starting from some $\grpu{3}$ irrep with some lowest number of
oscillator quanta, the \textit{lowest grade irrep} (LGI).  Laddering with the
symplectic raising operator ($A\propto b^\dagger b^\dagger$), which creates
pairs of oscillator quanta, generates the rest of the symplectic irrep, which is
an infinite tower of $\grpu{3}$ irreps of increasing oscillator number.  In
practice, for SpNCCI calculations, this irrep must be truncated at some finite
number of oscillator quanta.

To understand the quantum numbers for the SpNCCI basis, we first note that the
group $\grpu{3}=\grpu{1}\times\grpsu{3}$ is the product of the $\grpu{1}$
generated by the harmonic oscillator number operator, with quantum number $N$,
and the Elliott $\grpsu{3}$ group, which provides quantum numbers
$(\lambda,\mu)$.  Thus, a $\grpu{3}$ irrep is labeled by quantum numbers
$\omega\equiv N_\omega(\lambda_\omega,\mu_\omega)$.  The structure of a
symplectic irrep is uniquely defined by the $\grpu{3}$ quantum numbers
$\sigma\equiv N_\sigma(\lambda_\sigma,\mu_\sigma)$ of its LGI, which thus serve
as the $\grpsptr$ quantum numbers of the symplectic irrep.  The detailed
structure of an $\grpsptr$ irrep in terms of $\grpu{3}$ irreps is illustrated in
the inset of Fig.~\ref{fig-spncci-reorganization}, specifically for the
$16(2,1)$ irrep of $\isotope[8]{Be}$.\footnote{The $\grpu{1}$ label $N$
  appearing in the $\grpu{3}$ labels $N(\lambda,\mu)$ is actually not the number
  of oscillator quanta, \textit{per se}, but rather the dimensionless oscillator
  Hamiltonian, which also includes a zero-point contribution of $3/2$ for each
  nucleon~\cite{rowe1985:micro-collective-sp6r}.  Thus, in the
  lowest Pauli-allowed oscillator configuration of $\isotope[8]{Be}$, the four
  $s$-shell nucleons contribute $3/2$ each, and the four $p$-shell nucleons
  contribute $5/2$ each, giving $N=16$, as in the
  $16(2,1)$ irrep shown.}

At the bottom of the group chain, $\grpso{3}$ is the orbital angular momentum
group, which provides the quantum number $L$.  While the $\grpsptr$ subgroup
chain describes orbital (spatial) structure, there is also the complementary
$\grpsu{2}$ spin group, with quantum number $S$.  Orbital and spin angular
momenta couple to give total angular momentum $J$.  The SpNCCI basis is
therefore classified according to symmetry labels for the subgroup chain
\begin{equation}
\label{eqn-sp-chain}
[\underset{\mathclap{N_\sigma(\lambda_\sigma,\mu_\sigma)}}{\grpsptr}
\supset
\underset{\mathclap{N_\omega(\lambda_\omega,\mu_\omega)}}{\grpu{3}}
\supset
\underset{L}{\grpso{3}}
]\times
\underset{S}{\grpsu[S]{2}}
\supset
\underset{J}{\grpsu[J]{2}}.
\end{equation}
Recall that here $N_\omega(\lambda_\omega,\mu_\omega)$ are the $\grpu{3}$
quantum number of the basis state itself~--- so, if we are only interested in
$\grpu{3}$ symmetry, we can simply designate these $N(\lambda,\mu)$. Then,
$N_\sigma(\lambda_\sigma,\mu_\sigma)$ are the $\grpu{3}$ quantum numbers of the
LGI from which the entire symplectic irrep is built.

When defining an NCCI basis, it is more transparent and convenient to work with
the number of oscillator \textit{excitations} $\Nex$ relative to the lowest
Pauli-allowed oscillator configuration for the nucleus.  Configurations with
$\Nex=0$ are traditionally known as ``$0\hw$'' configurations, those with
$\Nex=2$ as ``$2\hw$'' configurations, \textit{etc.}\ Similarly, for a SpNCCI
basis, we shall refer to the number $\Nwex$ (or simply $\Nex$) of excitation
quanta for $\grpu{3}$ irreps and $\Nsex$ for the LGIs of $\grpsptr$ irreps.
Thus, in the symplectic decomposition of the many-body space shown schematically
Fig.~\ref{fig-spncci-reorganization} (right), each disk (light shading)
respresents states with a given $\Nex=0$, $2$, or $4$, while each cone (dark
shading) represents a symplectic irrep with LGI at $\Nsex=0$, $2$, or $4$.

In our SpNCCI framework, matrix
elements of the Hamiltonian (or other operators) in the symplectic many-body
basis are determined from certain ``seed'' matrix elements by use of symplectic
laddering operations.  The approach is based on ideas initially proposed by
Suzuki and
Hecht~\cite{suzuki1986:sp6r-alpha-cluster-me,suzuki1986:sp6r-cluster}, but these
have now been extended to accommodate general, realistic internucleon
interactions.

In order to evaluate the matrix elements of an operator in the SpNCCI basis, we
first decompose the operator in terms of $\grpsu{3}$-coupled components, or
\textit{unit tensors}.  We then apply commutation relations of these unit
tensors with the symplectic laddering operators to recursively express all
matrix elements in terms of matrix elements among certain lowest states,
specifically, the LGIs.  We explicitly expand these LGIs in terms of the
$\grpsu{3}$ no-core shell model [$\grpsu{3}$-NCSM] basis and use the
$\grpsu{3}$-NCSM code \texttt{LSU3shell}~\cite{dytrych2016:su3ncsm-12c-efficacy}
to evaluate the seed matrix elements.  The recurrence then involves coefficients
obtained through vector coherent state (VCS)
methods~\cite{rowe1983:sp6r-me,rowe1984:sp6r-vcs,hecht1987:vcs}.

\section{Illustration: Symplectic symmetry in \boldmath$\isotope[6]{Li}$}
\setcounter{equation}{0}
\label{sec-symmetry}

\begin{figure}[t]
\begin{center}
\includegraphics[width=0.55\hsize]{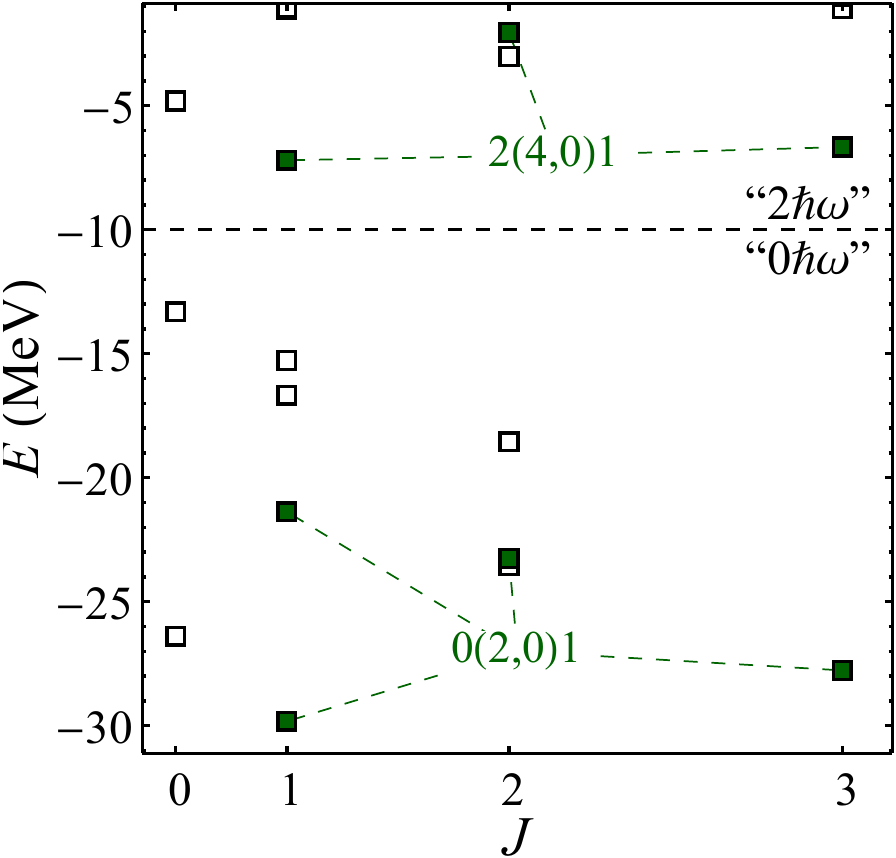}
\caption{Energy eigenvalues from an SpNCCI calculation for
  $\isotope[6]{Li}$.  Energies are
  plotted \textit{vs.}\ $J(J+1)$, as conventional for rotational
  analysis~\cite{maris2015:berotor2}.  States below the dashed line are
  predominantly $0\hw$, while those above the line have predominantly $2\hw$ (or
  higher) contributions.  States having dominant $\grpu{3}\times\grpsu{2}$
  contributions with $\Nex(\lambda,\mu)S$ quantum numbers $0(2,0)1$ and $2(4,0)1$
  are highlighted (solid symbols, as labeled).}
\label{fig-6li-jj1}
\end{center}
\figurecompensation
\end{figure}
\begin{figure}[t]
\begin{center}
\includegraphics[width=0.75\hsize]{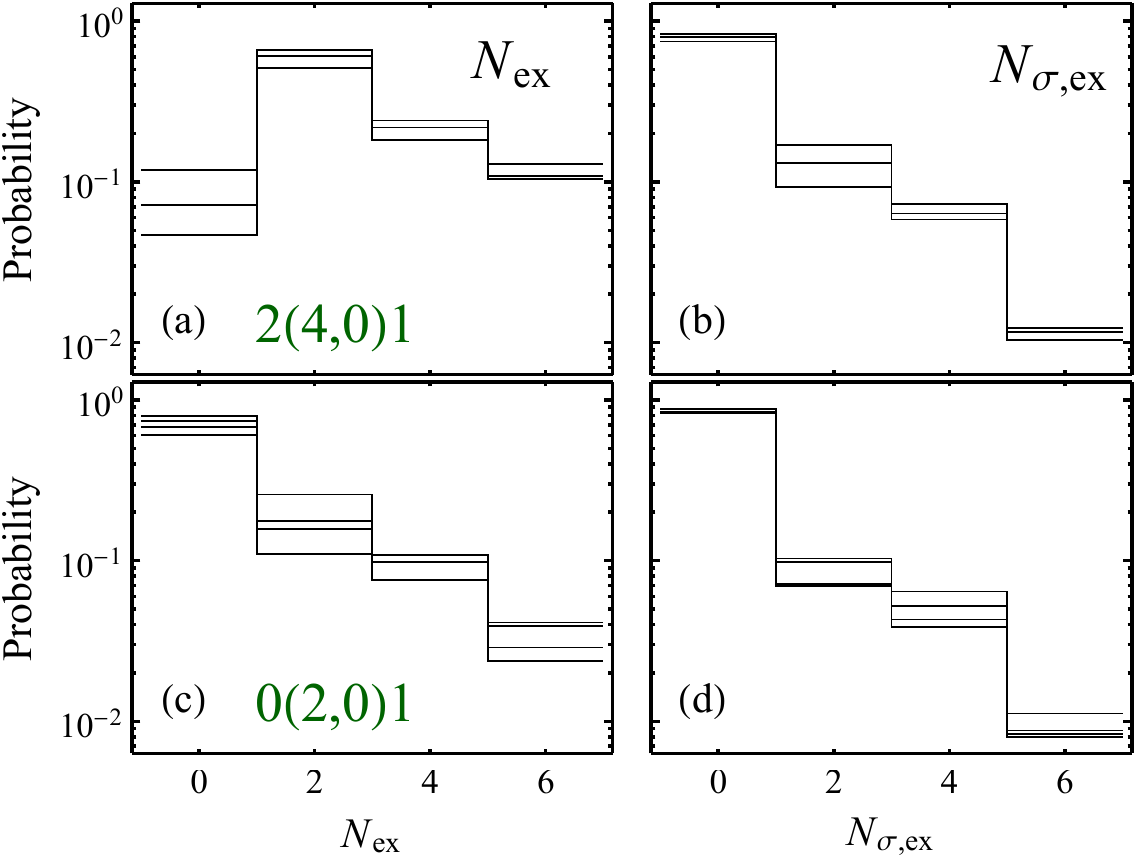}
\caption{Decompositions of calculated eigenstates by oscillator excitation quanta
  $\Nex$ (left) and by the $\Nsex$ of the contributing symplectic irrep (right).
  Decompositions are shown for the highlighted states from
  Fig.~\ref{fig-6li-jj1}, with dominant
  contributions from $\grpu{3}\times\grpsu{2}$ symmetry $0(2,0)1$ (bottom) or
  $2(4,0)1$ (top).  Only oscillator
  configurations with even values of $\Nex$ contribute to natural parity states.}
\label{fig-6li-Nex}
\end{center}
\figurecompensation
\end{figure}

Since SpNCCI calculations are carried out in an $\grpsptr\supset\grpu{3}$ basis,
we can immediately extract the decomposition of each eigenfunction into
contributions with different symmetry characters, for all the groups in the
chain~(\ref{eqn-sp-chain}).  That is, we can decompose the wave function
according to any combination of the quantum numbers
$\sigma=N_\sigma(\lambda_\sigma,\mu_\sigma)$ for $\grpsptr$ and
$\omega=N_\omega(\lambda_\omega,\mu_\omega)$ for $\grpu{3}$, as well as the more
familiar orbital angular momentum $L$ and spin $S$.

By examining such
decompositions, we can start to answer some of the questions we laid out above:
Do the highly-excited oscillator-basis contributions to the nuclear wave
functions primarily come from low-lying symplectic irreps, suggesting that
symplectic truncation may be feasible?  And can an $\grpsptr\supset\grpu{3}$
irrep structure provide a useful classification scheme for understanding the
nuclear eigenspectrum?

Here we consider a SpNCCI calculation of $\isotope[6]{Li}$, for which the
low-lying energy spectrum is shown in Fig.~\ref{fig-6li-jj1}.  This calculation
is based on the JISP16 internucleon interaction~\cite{shirokov2007:nn-jisp16},
with no Coulomb interaction (\textit{i.e.}, pure isoscalar).  For initial
examination and benchmarking purposes, we simply choose a space which is
truncated at $6$ oscillator quanta, with oscillator basis parameter
$\hw=20\,\MeV$.  That is, we take all symplectic irreps with LGIs having up to
$\Nsmax=6$ excitation quanta, then truncate each symplectic irrep to basis
states with up to $\Nmax=6$ excitation quanta.\footnote{The resulting spectrum
  is identical to that for a traditional $\Nmax=6$ $M$-scheme NCCI calculation,
  since both spaces include all intrinsic excitations of up to $6$ quanta (this
  equivalence permits rigorous numerical validatation against traditional NCCI
  codes such as
  MFDn~\cite{maris2010:ncsm-mfdn-iccs10,aktulga2013:mfdn-scalability}).  While
  the traditional $M$-scheme NCCI basis consists of laboratory-frame Slater
  determinants (which include center-of-mass excitations and which, while having
  definite $M$, are in general admixtures of all $J\geq\abs{M}$), the SpNCCI
  basis is defined in the intrinsic frame (consisting of states which are free
  of center-of-mass excitations and which are also naturally $J$-coupled).
  Compare the resulting dimension of $197\,822$ for the laboratory-frame $M=0$
  space~\cite{maris2009:ncfc} with $3484$ for the intrinsic-frame $J=0$
  space~\cite{luo2013:su3cmf}, for $\isotope[6]{Li}$ at $\Nmax=6$.}

Each of the states in Fig.~\ref{fig-6li-jj1} is found to have at most one or two
dominant $\grpu{3}\times\grpsu{2}$ contributions, that is, well-defined $\omega$
and spin quantum numbers.  However, we focus on two specific
$\grpu{3}\times\grpsu{2}$ ``families'' of states in this discussion, identified
by the filled symbols in Fig.~\ref{fig-6li-jj1}.  We shall see that these together
form part of a larger family of states with the same dominant symplectic
symmetry.

Let us start by classifying the states in the spectrum of Fig.~\ref{fig-6li-jj1}
according to contributions from configurations with different numbers of
oscillator quanta.  The calculated states
below an excitation energy of $\sim20\,\MeV$, as demarcated by the horizontal
dashed line in Fig.~\ref{fig-6li-jj1}, are dominated by $\Nex=0$ contributions.
That is, in traditional shell model parlance, they have predominantly $0\hw$
character.  Above this point in the spectrum, eigenstates dominated by $\Nex=2$
contributions, or $2\hw$ states, begin to appear.

To provide a more quantitative picture of this situation, decompositions by
$\Nex$ are shown in Fig.~\ref{fig-6li-Nex}~(left): specifically, decompositions for the $0\hw$ states
highlighted at the bottom of Fig.~\ref{fig-6li-jj1} are overlaid in the lower
panel [Fig.~\ref{fig-6li-Nex}(b)], and decompositions for the $2\hw$ states
highlighted at the top of Fig.~\ref{fig-6li-jj1} are overlaid in the upper
panel [Fig.~\ref{fig-6li-Nex}(a)].\footnote{When interpreting these, and
  subsequent, decompositions in an oscillator basis, it is important to keep in
  mind that they reflect a particular choice of oscillator length (here,
  corresponding to $\hw=20\,\MeV$) and are subject to change as the oscillator
  basis is dilated to other values of $\hw$.  Therefore, one should be cautious
  in attaching undue physical significance, \textit{e.g.}, interpreting them to
  mean that any given state has fundamentally a ``$0\hw$'' or ``$2\hw$'' nature
  with respect to some true mean-field vacuum.}  Clearly the ``$0\hw$''
states are not described entirely in the valence shell, but rather are heavily
``dressed'' with excited oscillator configurations, nor are the ``$2\hw$'' states
entirely orthogonal to the valence space.  However, the $0\hw$ states are
dominated by $\Nex=0$ contributions at the $60\%$--$80\%$ level, while, for the
$2\hw$ states, the $\Nex=0$ contributions fall below the $15\%$ level.

\begin{figure}[t]
\begin{center}
\includegraphics[width=0.97\hsize]{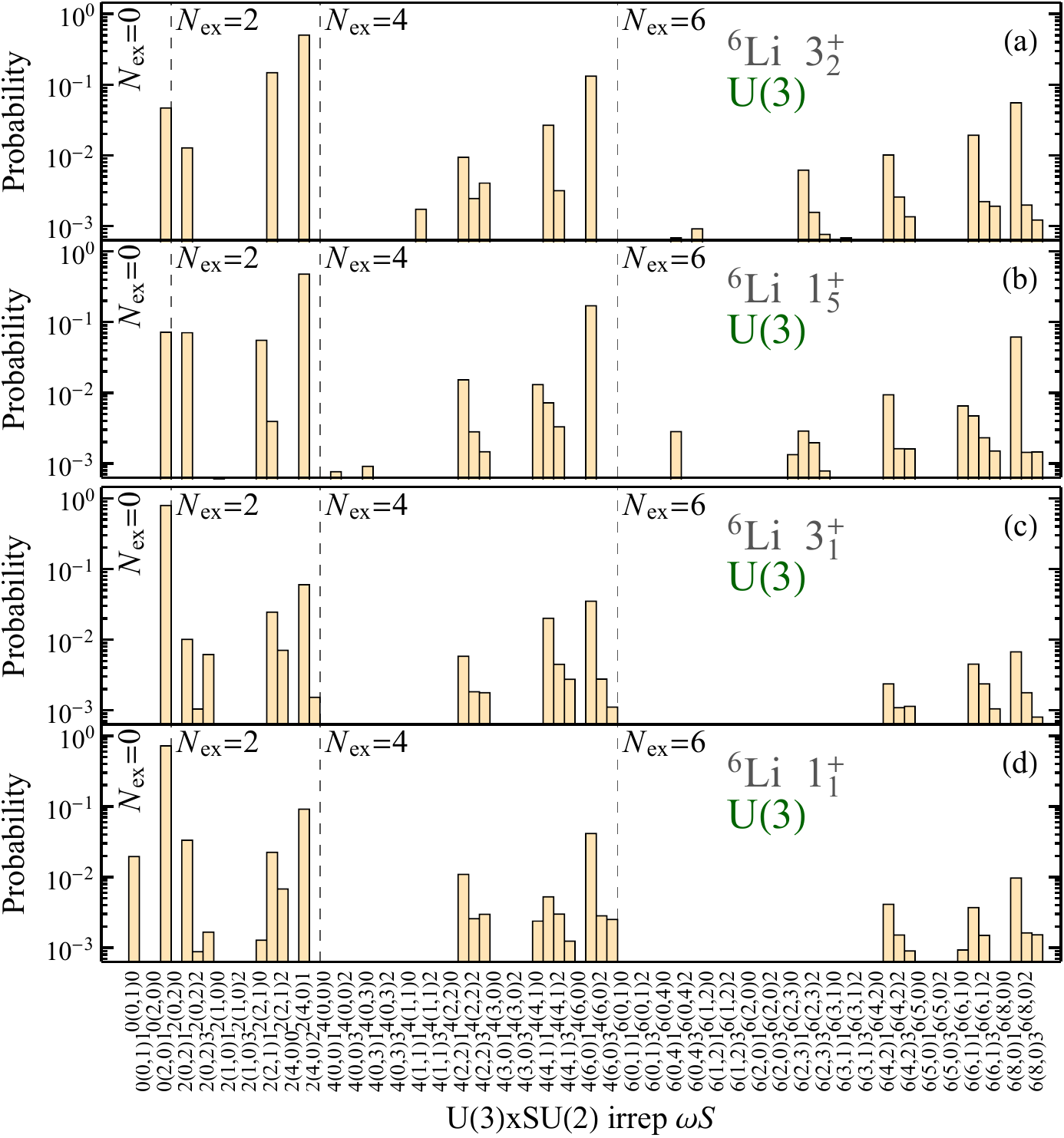}
\caption{Decompositions of calculated eigenstates by $\grpu{3}\times\grpsu{2}$
  contributions $\Nex(\lambda,\mu)S$: for the lowest $1^+$ and $3^+$ states~(c--d)
  and for excited $1^+$ and $3^+$ states of predominantly $2\hw$ character~(a--b).
The contributions are ordered, from left to right in the figure, first by
$\Nex$, and then by $\grpsu{3}$ and spin quantum numbers.  
}
\label{fig-6li-u3su2}
\end{center}
\figurecompensation
\end{figure}

We now look at the full decomposition with respect to $\grpu{3}\times\grpsu{2}$
character, labeled by the $\grpu{3}\times\grpsu{2}$ quantum numbers
$\Nex(\lambda,\mu)S$, in Fig.~\ref{fig-6li-u3su2}.  The states in the $0\hw$
family highlighted at the bottom of Fig.~\ref{fig-6li-jj1} all have dominant
$\grpu{3}\times\grpsu{2}$ contribution $0(2,0)1$.  [For the ground state of
  $\isotope[6]{Li}$, the $\grpu{3}$ structure~--- and $0(2,0)1$ dominance~---
  was explored by Dytrych \textit{et al.}~\cite{dytrych2013:su3ncsm}, on the
  basis of \textit{ab initio} $\grpsu{3}$-NCSM calculations.] Contributions from
the other $0\hw$ $\grpu{3}\times\grpsu{2}$ irreps appear at the few percent
level.  This pattern of $\grpu{3}\times\grpsu{2}$ contributions is shown for the
$1^+_1$ ground state and $3^+_1$ first excited state in
Fig.~\ref{fig-6li-u3su2}(c,d).  The states in the $2\hw$ family highlighted at
the top of Fig.~\ref{fig-6li-jj1} instead have dominant
$\grpu{3}\times\grpsu{2}$ contribution $2(4,0)1$, as shown for the $1^+$ and
$3^+$ states of that family in Fig.~\ref{fig-6li-u3su2}(a,b).
\begin{figure}[t]
\begin{center}
\includegraphics[width=0.97\hsize]{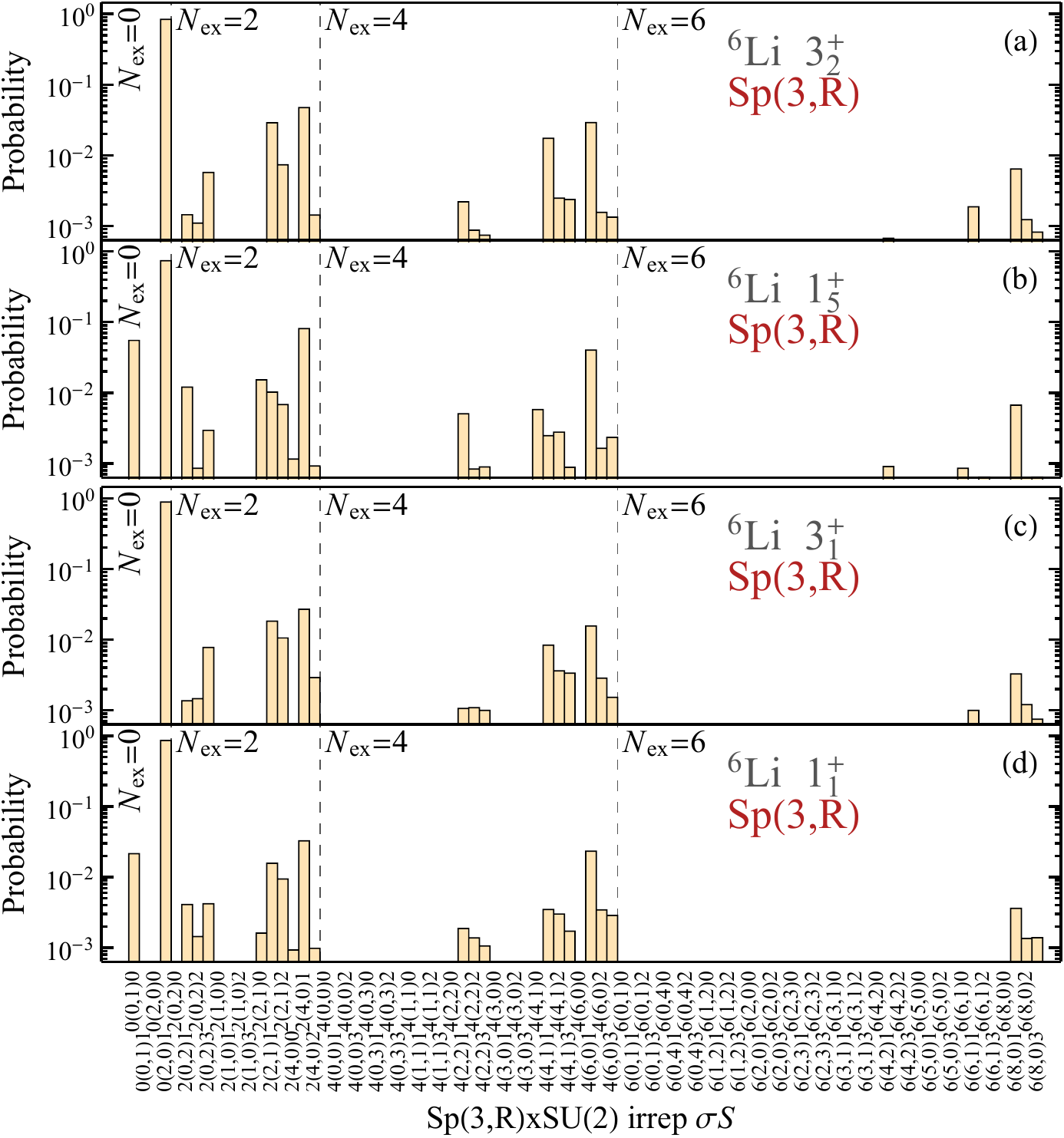}
\caption{Decompositions of calculated eigenstates by $\grpsptr\times\grpsu{2}$
  contributions $\Nsex(\lambda_\sigma,\mu_\sigma)S$: for the lowest $1^+$ and $3^+$ states~(c--d)
  and for excited $1^+$ and $3^+$ states of predominantly $2\hw$ character~(a--b).}
\label{fig-6li-sp3rsu2}
\end{center}
\figurecompensation
\end{figure}

While the $\grpu{3}\times\grpsu{2}$ decomposition begins to tell us about the
symmetry properties of these states, it provides only partial information on the
symplectic structure.  Any SpNCCI basis state with $\Nex$ excitation
quanta comes from a symplectic irrep with $\Nsex\leq\Nex$.  That is, the basis
state may itself be part of an LGI ($\Nsex=\Nex$), or it may be obtained by the
action of the symplectic raising operator from a lower LGI ($\Nsex<\Nex$).

Each
bar in the $\grpu{3}\times\grpsu{2}$ decomposition histograms in
Fig.~\ref{fig-6li-u3su2} represents the total probability contribution of many
basis states from several $\grpu{3}\times\grpsu{2}$ irreps sharing the same
quantum numbers.  For instance, there are in fact $30$ different
$\grpu{3}\times\grpsu{2}$ irreps with quantum numbers $2(4,0)1$, for
$\isotope[6]{Li}$.  One of these comes from the $\Nsex=0$ symplectic irrep
$0(2,0)1$, while the rest are themselves LGIs of $\Nsex=2$ symplectic
irreps.

Before breaking each wave function down into contributions from specific
$\grpsptr$ quantum numbers, it is informative to simply look at the
decomposition of each wave function with respect to how ``excited'' the
contributing symplectic irreps are.  Decompositions by $\Nsex$ are shown in
Fig.~\ref{fig-6li-Nex}~(right).

Since the $\Nsex=0$ contribution must be at least as large as the $\Nex=0$
contribution, clearly the $0\hw$ states will have a significant $\Nsex=0$
contribution.  In particular, for the highlighted family of states with
$\grpu{3}\times\grpsu{2}$ character $0(2,0)1$ [Fig.~\ref{fig-6li-Nex}(d)], the
$\Nsex=0$ contributions are in the $70\%$--$90\%$ range.  Thus, a substantial portion of
the excited oscillator contributions [Fig.~\ref{fig-6li-Nex}(c)] actually
comes from $\Nsex=0$ sympletic irreps.  This is encouraging for the viability of symplectic
truncation schemes.

But what about the $2\hw$ states?  For the highlighted family of states with
$\grpu{3}\times\grpsu{2}$ character $2(4,0)1$ [Fig.~\ref{fig-6li-Nex}(b)],
recall that this $2(4,0)1$ contribution could come either from the $\Nsex=0$
symplectic irrep $0(2,0)1$ or from the $2(4,0)1$ LGIs of $\Nsex=2$ symplectic
irreps.  So, are these $2\hw$ states dominated by contributions from $\Nsex=0$
symplectic irreps or from $\Nsex=2$ symplectic LGIs?  From the $\Nsex$
decompositions [Fig.~\ref{fig-6li-Nex}(b)], it is immediately
apparent that they are dominated by contributions from $\Nsex=0$, at the $>70\%$
level.

The full decompositions with respect to $\grpsptr\times\grpsu{2}$ character,
labeled by the $\grpsptr\times\grpsu{2}$ quantum numbers
$\Nsex(\lambda_\sigma,\mu_\sigma)S$, are shown in Fig.~\ref{fig-6li-sp3rsu2}.
The most notable feature apparent in Fig.~\ref{fig-6li-sp3rsu2} is the dominance
of the contribution from the $0(2,0)1$ symplectic irrep, for all these states.
In fact, all the highlighted states in Fig.~\ref{fig-6li-jj1}~--- both the
low-lying $0\hw$ states with $0(2,0)1$ $\grpu{3}\times\grpsu{2}$ character and
the high-lying $2\hw$ states with $2(4,0)1$ $\grpu{3}\times\grpsu{2}$
character~--- receive their dominant contribution from this single $\Nsex=0$
symplectic irrep, the $0(2,0)1$ symplectic irrep.

Thus, states with very different oscillator excitation content
[Fig.~\ref{fig-6li-Nex} (left)], and consequently very different $\grpu{3}$
content (Fig.~\ref{fig-6li-u3su2}), nonetheless form a larger family of states
sharing the same symplectic symmetry.  This is fundamentally the idea of the
classification scheme for the SpNCCI \textit{basis} states in an
$\grpsptr\supset\grpu{3}$ scheme~(\ref{eqn-sp-chain}): many $\grpu{3}$ irreps
form a single $\grpsptr$ irrep, as illustrated in
Fig.~\ref{fig-spncci-reorganization} (inset).  However, here we find the same
$\grpsptr\supset\grpu{3}$ organizational scheme holding for the physical spectrum of energy \textit{eigen}states
obtained from an \textit{ab initio} calculation.

\section{Conclusions}
\setcounter{equation}{0}

In conclusion, a framework for \textit{ab initio} no-core configuration
interaction calculations in an $\grpsptr\supset\grpu{3}$ symplectic basis is now
in place.  This scheme builds on an existing $\grpsu{3}$-coupled framework for
the nuclear problem and combines it with the group-theoretical machinery for
$\grpsptr$.

In initial calculations, illustrated here with $\isotope[6]{Li}$, we confirm
$\grpsptr$ as an approximate symmetry of states throughout the low-energy
spectrum.  These states are characterized by mixing of a few dominant symplectic
irreps.  Then, families of states arise with similar symplectic structure,
despite their differing $\grpu{3}$ content.

To take full advantage of approximate symplectic symmetry in nuclei, as a means of accelerating
convergence of NCCI calculations, it will be necessary to go beyond the simple
benchmark $\Nmax$ truncation scheme for the SpNCCI space illustrated here.
While some gains can likely be made by truncating the space to dominant
$N_\sigma(\lambda_\sigma,\mu_\sigma)S$ symplectic subspaces, these subspaces are
highly degenerate, especially as we go to higher $\Nsex$.

The most effective
truncation will therefore likely come by identifying the few dominant symplectic
irreps from within these subspaces (\textit{e.g.}, by some variant of importance
truncation~\cite{roth2007:it-ncsm-40ca}).  Once these ``Hamiltonian preferred''
symplectic irreps are identified, a more stringent truncation can be
carried out to a relatively small number of symplectic irreps within each subspace.

\clearpage 
\section*{Acknowledgements}

We thank David J.~Rowe for invaluable assistance with the $\grpsptr$ formalism
and Chao Yang, Pieter Maris, Calvin W.~Johnson, and Patrick J.~Fasano for
discussions of the computational implementation.

This material is based upon work supported by the U.S.~Department of Energy,
Office of Science, Office of Nuclear Physics, under Award Number
DE-FG02-95ER-40934, by the U.S.~Department of Energy, Office of Science, Office
of Workforce Development for Teachers and Scientists, Graduate Student Research
(SCGSR) program, under Contract Number DE-AC05-06OR23100, by the Research
Corporation for Science Advancement, under a Cottrell Scholar Award, and by the
Czech Science Foundation, under Grant Number 16-16772S.  TRIUMF receives federal
funding via a contribution agreement with the National Research Council of
Canada.

This research used computational resources of the University of Notre Dame
Center for Research Computing and of the National Energy Research Scientific
Computing Center (NERSC), a U.S.~Department of Energy, Office of Science, user
facility supported under Contract~DE-AC02-05CH11231.

\providecommand{\APSLONG}{}
\providecommand{\ELSEVIER}{}


\enddocument